\documentclass[%
 manuscript,%
 amssymb, amsmath,%
 aip,cha,%
]{revtex4-1}

\usepackage{float}
\usepackage{docs}%
\usepackage{bm}%
\usepackage[colorlinks=true,linkcolor=blue]{hyperref}%
\usepackage{graphicx}
\usepackage{epstopdf}
\expandafter\ifx\csname package@font\endcsname\relax\else
 \expandafter\expandafter
 \expandafter\usepackage
 \expandafter\expandafter
 \expandafter{\csname package@font\endcsname}%
\fi
\begin{document}

\title{An Environment-dependent Semi-Empirical Tight Binding Model Suitable for Electron Transport in Bulk Metals, Metal Alloys, Metallic Interfaces and Metallic Nanostructures II - Application - Effect of Quantum Confinement and Homogeneous Strain on Cu Conductance}%

\author{Ganesh Hegde}%
\email{ghegde@purdue.edu}
\affiliation{Network for Computational Nanotechnology (NCN),\\Department of Electrical and Computer Engineering, Purdue University\\West Lafayette, Indiana 47907, USA}%
\author{Michael Povolotskyi}
\affiliation{Network for Computational Nanotechnology (NCN),\\Department of Electrical and Computer Engineering, Purdue University\\West Lafayette, Indiana 47907, USA}%
\author{Tillmann Kubis}
\affiliation{Network for Computational Nanotechnology (NCN),\\Department of Electrical and Computer Engineering, Purdue University\\West Lafayette, Indiana 47907, USA}%
\author{James Charles}
\affiliation{Network for Computational Nanotechnology (NCN),\\Department of Electrical and Computer Engineering, Purdue University\\West Lafayette, Indiana 47907, USA}%
\author{Gerhard Klimeck}
\email{gekco@purdue.edu}
\affiliation{Network for Computational Nanotechnology (NCN),\\Department of Electrical and Computer Engineering, Purdue University\\West Lafayette, Indiana 47907, USA}%
\date{\today}%
\begin{abstract}
The Semi-Empirical TB model developed in part I is applied to metal transport problems of current relevance in part II. A systematic study of the effect of quantum confinement, transport orientation and homogeneous strain on electronic transport properties of Cu is carried out. It is found that quantum confinement from bulk to nanowire boundary conditions leads to significant anisotropy in conductance of Cu along different transport orientations. Compressive homogeneous strain is found to reduce resistivity by increasing the density of conducting modes in Cu. The [110] transport orientation in Cu nanowires is found to be the most favorable for mitigating conductivity degradation since it shows least reduction in conductance with confinement and responds most favorably to compressive strain.
\end{abstract}
%\revised{August 2010}%
\maketitle
\tableofcontents
\section{Introduction}
Part I of this paper sequence discusses the shortcomings of semiclassical models from a predictive materials design standpoint. It also outlines some shortcomings of \textit{ab} \textit{initio} DFT methods in studying metal electronic structure problems at realistic length scales. As a scalable yet accurate and transferable alternative to DFT-based methods, a new Tight Binding (TB) model suitable for studying transport in metals is presented in part I.

Recent DFT-based studies have investigated issues like effect of surface roughness \cite{PhysRevB.79.155406}, liner layer \cite{PhysRevB.81.045406} and single grain boundary scattering in Cu \cite{feldman2010simulation}. Yet others have compared relative performance of Cu nanowires with Ag nanowires and carbon nanotubes \cite{zhou2008resistance, kharche2011comparative}

\begin{figure}[H]
	\centering
		\includegraphics[width=7.0in]{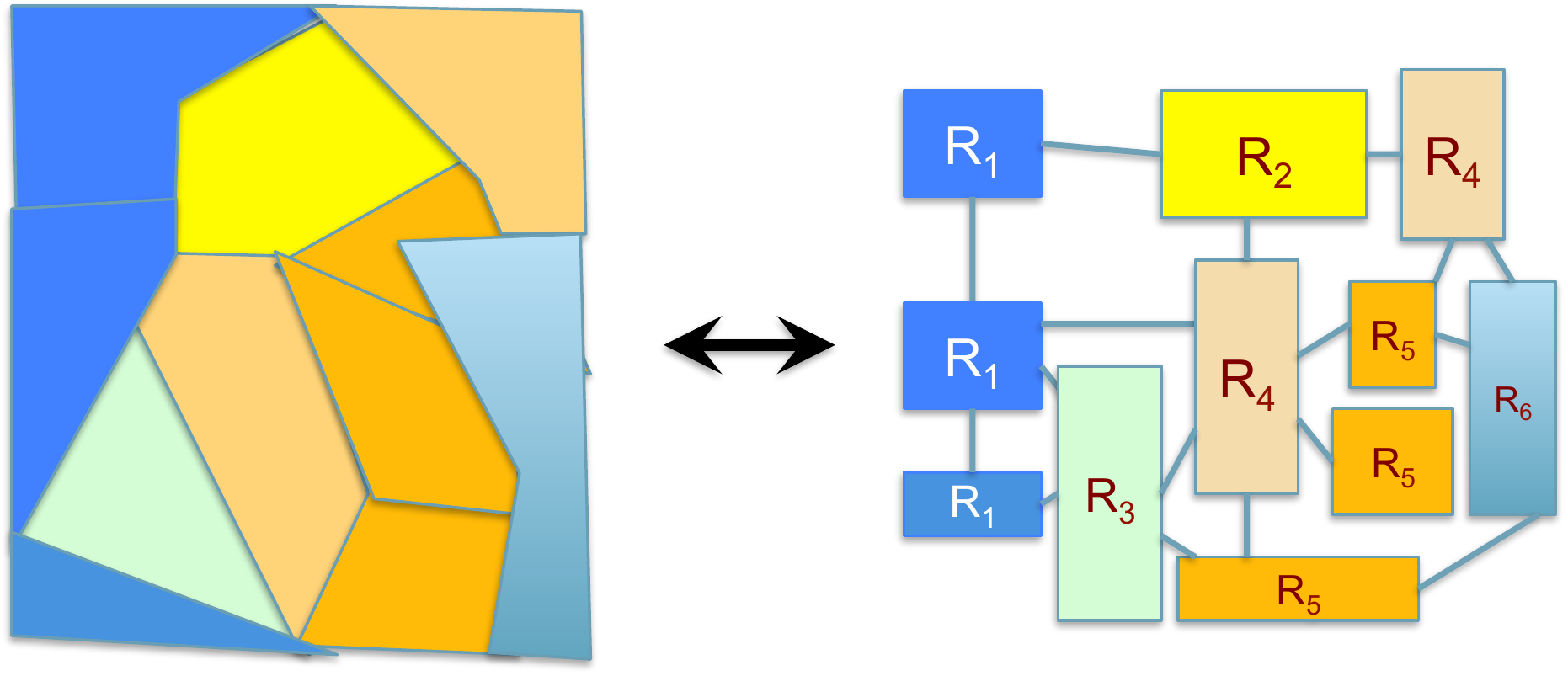}
	\caption{Schematic sketch of a polycrystalline Cu structure. Different colors indicate different crystalline orientations perpendicular to the plane of the paper, similar to data obtained from Electron Backscatter Diffraction experiments. The complicated polycrystalline structure can be viewed as an equivalent resistive network with ballistic resistances per unit length R as shown on the right in the absence of other scattering mechanisms}	
	\label{fig:schematic_polycrystalline}
\end{figure}

Polycrystalline structures such as metal interconnects can be treated as a network of interconnected conductors as shown in figure \ref{fig:schematic_polycrystalline}. As grain dimensions change from micrometers to nanometers, effects such as quantum confinement and individual grain crystalline orientation can be expected to influence electronic transport properties. It is therefore important that a systematic investigation into the dependence of metal conductance on factors such as quantum confinement and transport orientation be carried. To our knowledge, no such study exists in the present literature.

If certain transport orientations show significant conductance degradation versus others under confinement, a study of orientation dependent transport properties can provide important insights into the design of conductance-degradation mitigation schemes in polycrystalline interconnects.

Although models investigating conductance degradation in metallic thin films and nanowires exist in the literature, very few studies have proposed mitigative solutions. Patents on Cu conductivity improvement using compressive strain have been awarded in recent times \cite{haverty2007reducing}. A thorough examination of the patent reveals that the physical mechanisms behind conductivity improvement have only been speculated upon. For instance, it is speculated that a reduction in the Density of States (DOS) available for electron scattering causes improvement in conductivity. 

In part II of this paper, a systematic, fully quantum mechanical study of effect of confinement, transport orientation and homogeneous strain on Cu conductance for structures having cross sectional areas of up to 25 nm$^2$ is carried out using the new TB model developed in part I. 

The analysis done in this work indicates that there exists a significant anisotropy in ballistic conductance of different Cu transport orientations as dimensionality changes from bulk to nanowire. Contrary to existing speculation \cite{haverty2007reducing}, it is found that homogeneous compressive strain \textit{increases} the density of conducting modes per unit volume in bulk and nanowire Cu and is responsible for conductivity improvement in Cu.

\section{Computational Method}
\begin{figure}[H]
	\centering
		\includegraphics[width=7.0in]{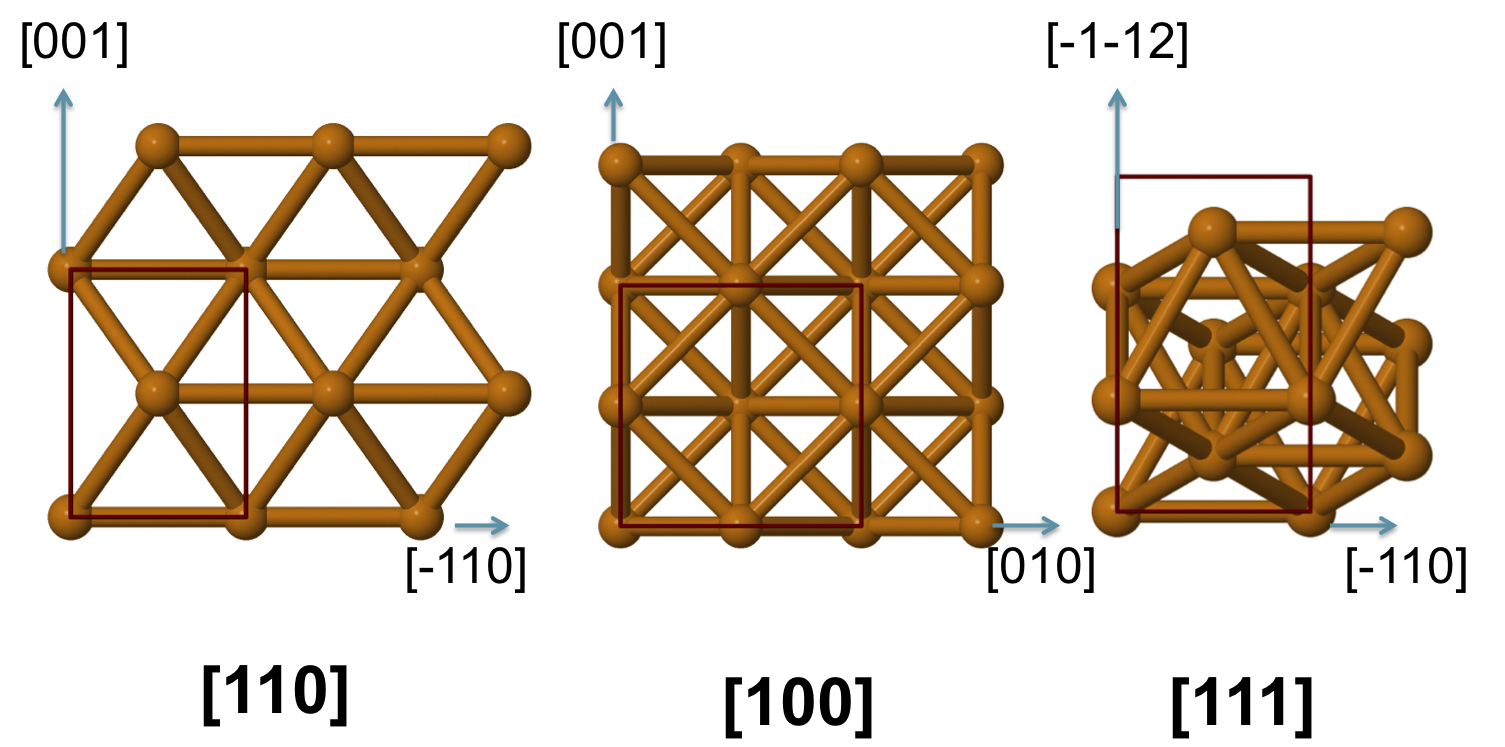}
	\caption{Square Cross sections of Cu oriented perpendicular to the direction of transport, which is indicated in bold. The box indicates the smallest cross sectional unit cell in the plane for the respective transport orientations}	
	\label{fig:CrossSectionsAll}
\end{figure}

To study effect of confinement, initially, ideal [100], [111] and [110] square Cu cross-sections of varying areas at the experimental zero strain lattice parameter $a_0 = 3.61 \AA$ are created in the Nanoelectronic Modeling Suite (NEMO5) \cite{fonseca2013efficient, steiger2011nanoelectronic} TB transport simulator as shown in figure \ref{fig:CrossSectionsAll}

These cross sections are then subjected to two different boundary conditions corresponding to two different confinement regimes - Bulk (3D periodic, no confinement) and nanowire (1D periodic, confinement along two dimensions) boundary conditions. The TB Hamiltonian is created using the Second Nearest Neighbor (2NN) parameters in appendix B of part I of this paper. The electronic band structure (E-$k$) is then obtained by solving for the eigenvalues of the TB Hamiltonian for each cross section for each boundary condition. From the band structure, the Fermi Level is computed by occupying  with electrons until the expected number of valence electrons is reached. For Cu, this number is 11(10 $d$ electrons and 1 $s$) valence electrons per atom so that overall charge neutrality is maintained. This condition is summarized in the following equation
\begin{equation}
\int{g(E) f(E) dE} = 11
\end{equation}
where $g(E)$ is the density of states per atom as a function of energy $E$, $f$ is the Fermi-Dirac occupation function.

From the electronic band structure, the number of conducting modes around Fermi Level can also be easily computed. This is nothing but the ballistic conductance for periodic structures \cite{datta1997electronic}.
\begin{equation}
\label{eq:ballisticconductance}
G = \frac{2e^2}{h}\sum\limits_{n}\int{T_{n}(E)\frac{-\partial f}{\partial E} dE}
\end{equation}
where the sum is over all bands $n$ , $T(E)$ is the number of conducting modes at energy level $E$. The quantity $\frac{2e^2}{h}$ is the fundamental quantum of conductance. In bulk, the ballistic conductance is computed over all transverse modes and then averaged. Due to symmetry, doubling the cross sectional area of a 3D periodic bulk cross section simply doubles the number of conducting modes in it. Consequently, it is sufficient to compute ballistic resistance for the smallest cross section corresponding to a each transport orientation in bulk. The conductance for larger cross sections in bulk can simply be obtained by multiplying the conductance of this smallest cross section by the ratio of the area of the respective cross section to the smallest cross section.  
%and $k_x$ and $k_y$ represent the transverse wave vectors in bulk and the gamma point for nanowires.
%\item Temperature averaged DOS/eV at Fermi Level for periodic and non-periodic structures
%\begin{equation}
%N = \int{N(E)\frac{-\partial f}{\partial E} dE}
%\end{equation}

To study effect of homogeneous strain the lattice parameter is varied from $\mp$ 5\% of the experimental lattice parameter $a_0$. Strain is computed as $\epsilon = \frac{a-a_0}{a_0}$ where $a$ is the strained lattice parameter.
The ballistic conductance in bulk and nanowires is then computed for each strain using the same procedure outlined above.

In addition to ballistic conductance, the resistivity of the Cu structures is also approximated using a modified Landauer approach \cite{zhou2008resistance}. This is outlined in the appendix of part I of this paper and is reproduced in the main body of this part of the paper for convenience. If a mean free path $\lambda$ between scattering events exists, then the ballistic conductance $G$ computed in equation \ref{eq:ballisticconductance} above can be modified using the following expression \cite{lundstrom2009fundamentals}
\begin{equation}
\label{eq:Gvslambda}
G' = G\frac{\lambda}{\lambda+L} \Rightarrow R' = \frac{1}{G}\frac{\lambda+L}{\lambda}
\end{equation}  
Where $L$ is the length of the conductor. 

If a linear fit for the ballistic conductance versus cross sectional area (A) curve is estimated, then we have 
\begin{equation}
G = kA+c \approx kA
\end{equation}
where $k$ is the slope of the linear fit and $c$ is the y-axis intercept that can be neglected for large cross sectional areas. If the length of the conductor is much greater than the mean free path (40 nm in Cu \cite{zhou2008resistance}) as in the case of infinitely periodic nanowires, we have $L>>\lambda$ and we can write.
\begin{equation}
R' = \frac{1}{kA}\frac{L}{\lambda}
\end{equation}
Equating this to Ohm's Law $R = \rho\frac{L}{A}$ we get an approximate resistivity as follows,
\begin{equation}
\rho = \frac{1}{k\lambda}
\label{eq:rho_versus_slope}
\end{equation}
Thus, the resistivity computed in this approach is inversely proportional to the slope of the conductance versus cross sectional area linear fit. It is also independent of the cross sectional area.

\section{Results and Discussion}
\begin{figure}
	\centering
		\includegraphics[width=7.0in]{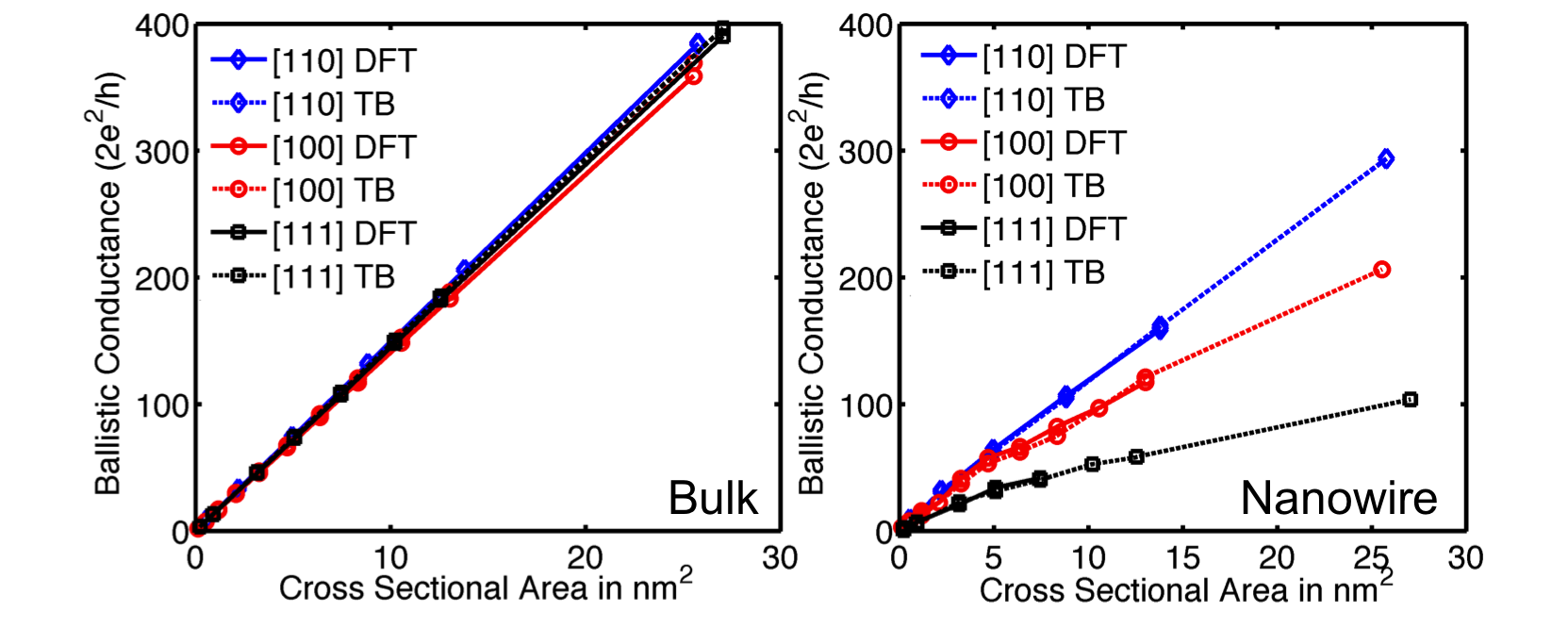}
	\caption{ Ballistic conductance versus cross sectional area for cross sections of different orientations exposed to bulk and nanowire boundary conditions. The bulk graph shows insignificant anisotropy while the nanowire graph shows significant anisotropy in ballsitic conductance versus orientation}	
	\label{fig:bulk_and_nanowire_transmission_versus_area}
\end{figure}

Figure \ref{fig:bulk_and_nanowire_transmission_versus_area} is a comparison of the ballistic conductance versus cross sectional area for bulk and nanowires for the zero strain lattice parameter $a_0 = 3.61 \AA$. In bulk, the ballistic conductance is largely isotropic versus orientation. It is well known that the Fermi surface of bulk, Face-Centered Cubic (FCC) Cu is largely spherical \cite{ashcroft1976solid}. Thus, the isotropy in conductance in different transport orientations can be expected. This is seen both in the DFT and TB results computed for bulk. The DFT results for bulk are included here as a means of validation in addition to results peresented in part I of this paper.

From figure \ref{fig:bulk_and_nanowire_transmission_versus_area}, it can also be seen that when Cu is quantum confined to form oriented nanowires, some important observations and inferences can be made. First, it can be seen that the ballistic conductance of nanowire cross sections is significantly smaller than their bulk counterparts. From this, it can be inferred that quantum confinement decreases the ballistic conductance of Cu significantly.

\begin{figure}
	\centering
		\includegraphics[width=7.0in]{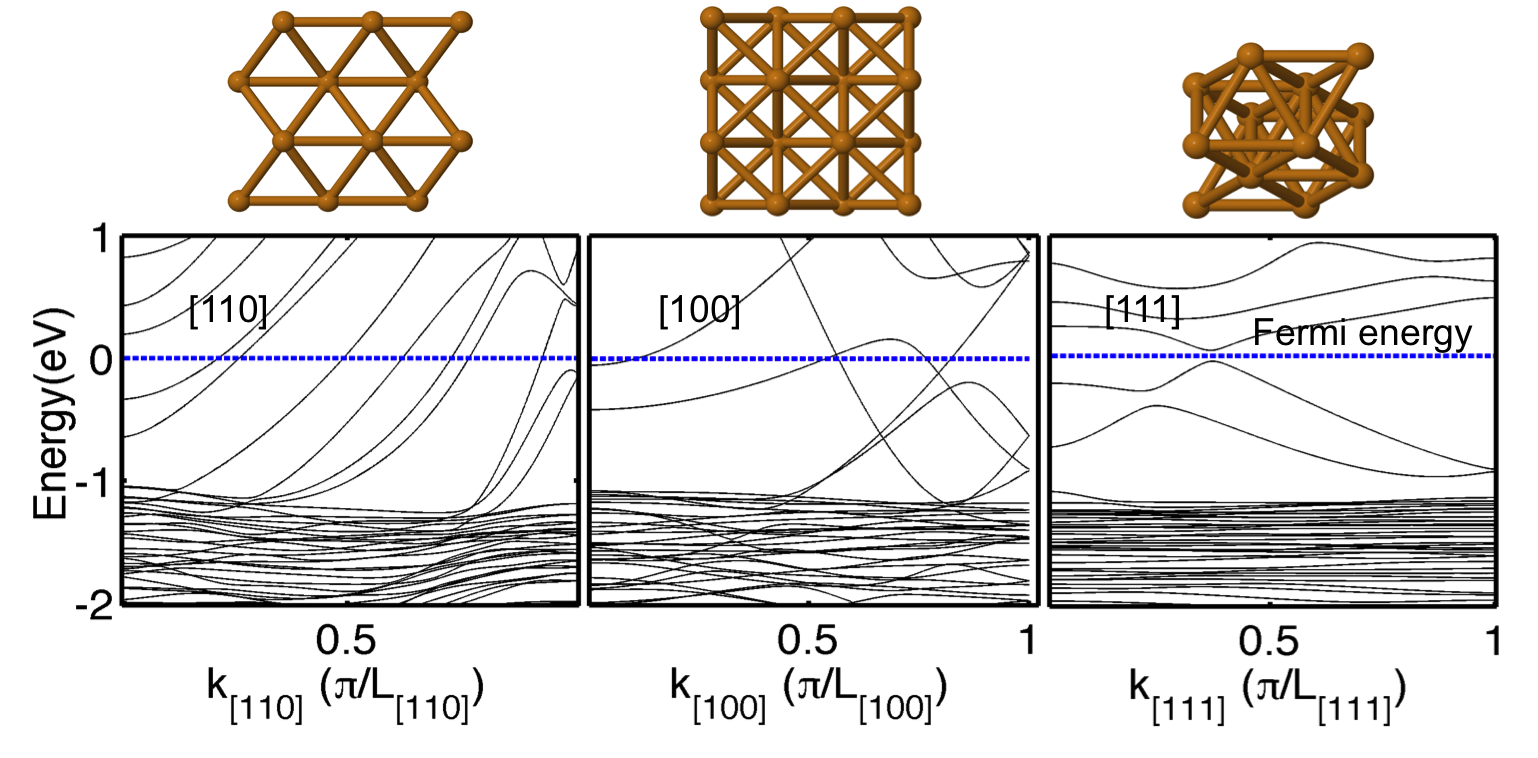}
	\caption{ Cross sectional atomic structures and band structures for square nanowires having (from left to right) [100], [110] and [111] orientation in direction of transport i.e. perpendicular to plane of paper. The structures shown are the smallest square cross sections of equivalent cross sectional area along different orientations.  It is evident that number of modes around Fermi Level for a 12 atom [110] unit cell is higher than a 16 atom [100] unit cell of similar area and 12 atom [111] unit cell}	
	\label{fig:wirebandstructures_and_structures}
\end{figure}

Secondly, the isotropy in conductance seen in bulk along different transport orientations is no longer seen in nanowires. Cu [110] nanowires have the largest ballistic conductance, followed by [100] and [111] oriented nanowires. This transport orientation-dependent anisotropy in ballistic conductance can be explained by examining nanowire band structures along different orientations. Figure \ref{fig:wirebandstructures_and_structures} shows the band structure of the smallest square cross sections of equalivalent cross sectional areas in the respective transport orientations. It is evident that the [110], [100] and [111] cross sections have very different band structures around the Fermi Level. Since the ballistic conductance is a measure of the number of conducting channels around the Fermi Level, this significant difference in band structures leads to significant conductance anisotropy in nanowires. As in the case of bulk, DFT results for small nanowire cross sections are also included as additional validation in figure \ref{fig:bulk_and_nanowire_transmission_versus_area}. It can be seen that the conductances computed in DFT match well qualitatively and quantitatively. It is possible to evaluate small nanowire cross sections of 10-15 nm$^2$ easily in DFT. Larger cross sections up to 25 nm$^{2}$, however, are computationally cumbersome for DFT and are hence computed using the new TB model. 

It can also be observed that conductance versus area relationship in figure \ref{fig:bulk_and_nanowire_transmission_versus_area} is linear in bulk. The relationship is non-linear for nanowires of small cross sections, especially of [111] orientation, but can be approximated by a linear fit for nanowires of large (i.e. $\geq$ 10-15 nm$^2$) cross sections. The linearity in ballistic conductance for large nanowire cross sections cannot be extrapolated accurately from DFT calculations for small nanowire cross sections. Full TB calculations of the band structure are thus necessary to justify the linearity assumption for larger nanowires.

Once it has been established that the conductance versus area relationship can be well approximated by a linear fit, resistivity can be computed according to the formalism outlined in equations \ref{eq:Gvslambda}-\ref{eq:rho_versus_slope}. Since the resistivity computed in this formalism is inversely proportional to slope of the conductance versus area curve, it can be inferred from the discussion above that simple quantum confinement increases resistivity by reducing the number of conducting channels available for conduction across all orientations as compared to bulk. From  figure \ref{fig:bulk_and_nanowire_transmission_versus_area}, it is also evident that the effect of confinement on resistivity is the least on [110] oriented nanowires while the most significant effect on [111] oriented nanowires.

\begin{figure}
	\centering
		\includegraphics[width=7.0in]{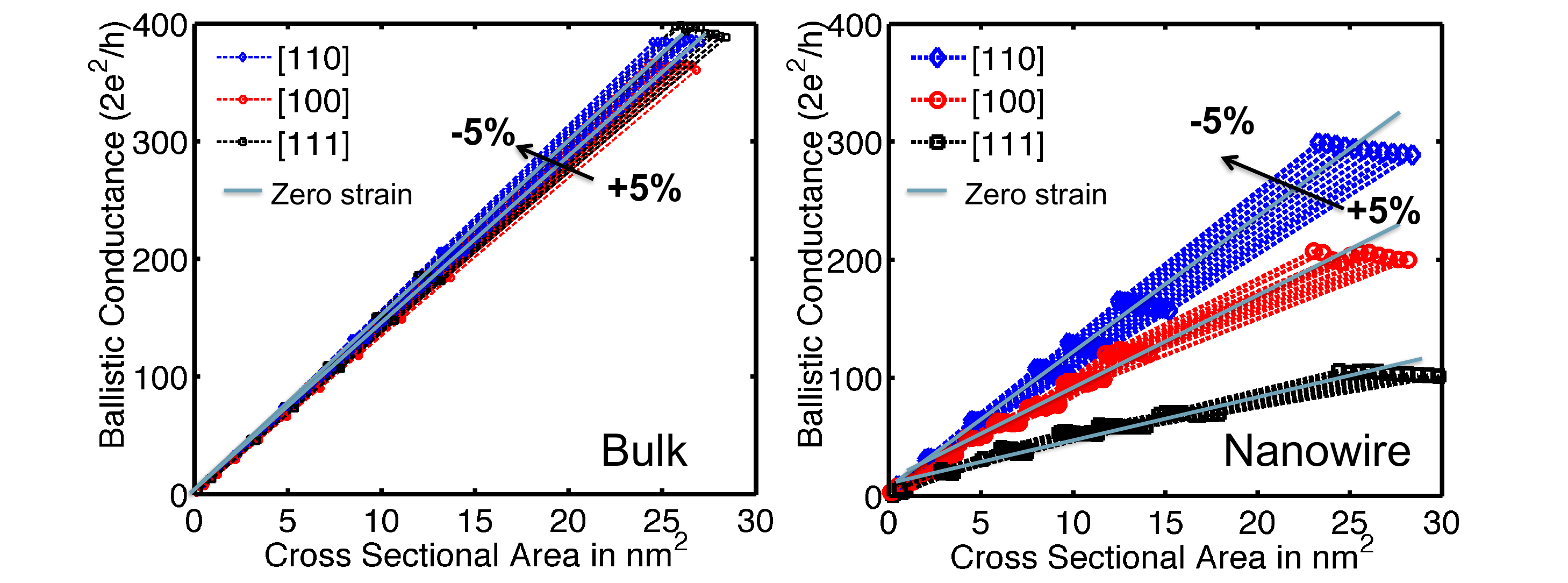}
	\caption{ Ballistic conductance versus cross sectional area for cross sections of different orientations exposed to bulk and nanowire boundary conditions versus strain. Also indicated schematically is the zero strain linear fit for each orientation}	
	\label{fig:bulk_and_nanowire_transmission_versus_area_strain}
\end{figure}

\begin{figure}
	\centering
		\includegraphics[width=7.0in]{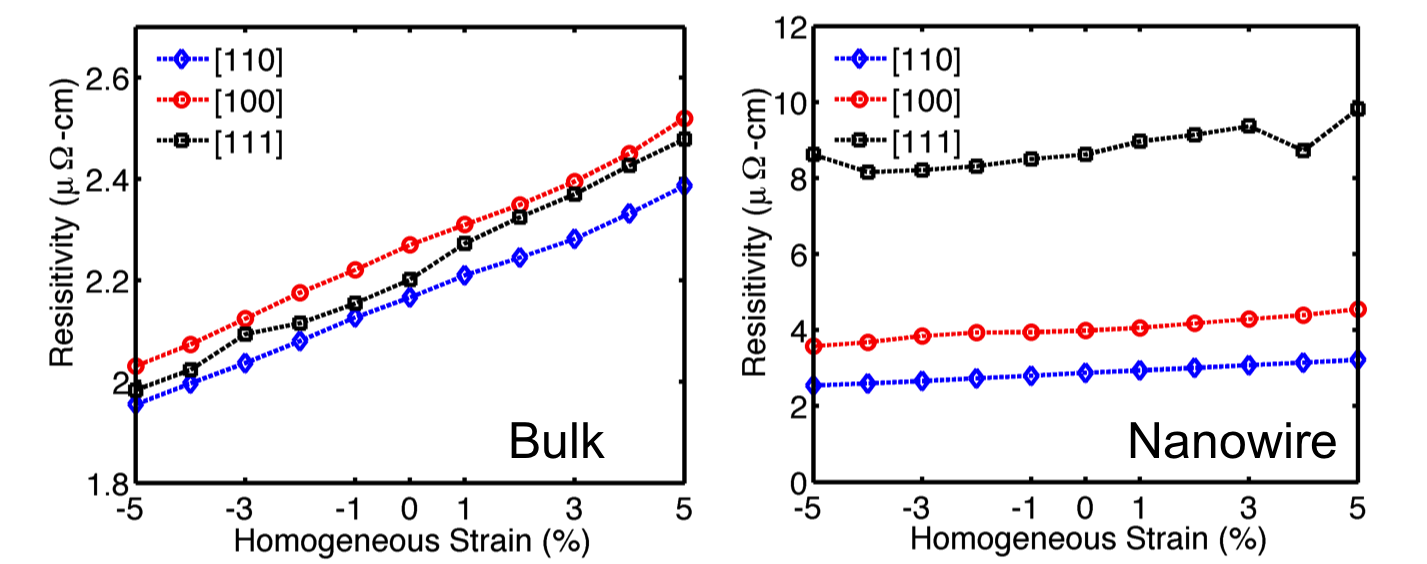}
	\caption{ Resistivity computed from the formula in equation \ref{eq:rho_versus_slope}. It is evident that the effect of confinement on resistivity is anisotropic. The effect of compressive strain is to reduce resistivity of Cu in bulk and nanowires.}
	\label{fig:resistivity_Strain_bulk_nanowire}
\end{figure}

Figure \ref{fig:bulk_and_nanowire_transmission_versus_area_strain} shows that homogeneous strain changes ballistic conductance significantly. Compressive strain increases ballistic conductance while tensile strain decreases it. In order to quantify the effect of strain on conductance, it is useful to extract a resistivity corresponding to each strain value for bulk and nanowire Cu according to equation \ref{eq:rho_versus_slope}. The results of such an analysis are summarized in figure \ref{fig:resistivity_Strain_bulk_nanowire}.

A linear estimate of the resistivity versus strain relationship in figure \ref{fig:resistivity_Strain_bulk_nanowire} indicates that compressive strain of 5\% decreases bulk Cu resistivity by 9 to 11\% as compared to the unstrained case. In nanowires, the effect of strain on resistivity seems more pronounced in the [110] and [100] orientation, where 5\% compressive strain decreases resistivity by 12\% and 11\% respectively, while the [111] shows a decrease of 7\% as compared to its unstrained value.

The physical mechanism responsible for the improvement of conductance and reduction in resistivity is quite evident from the conductance versus cross sectional area curves. Ballistic conductance is simply a measure of the number of conducting modes. Compressive strain increases the density of conducting modes per cross sectional area due to increased density of atoms per unit volume, thus increasing the ballistic conductance.

\begin{figure}
	\centering
		\includegraphics[width=7.0in]{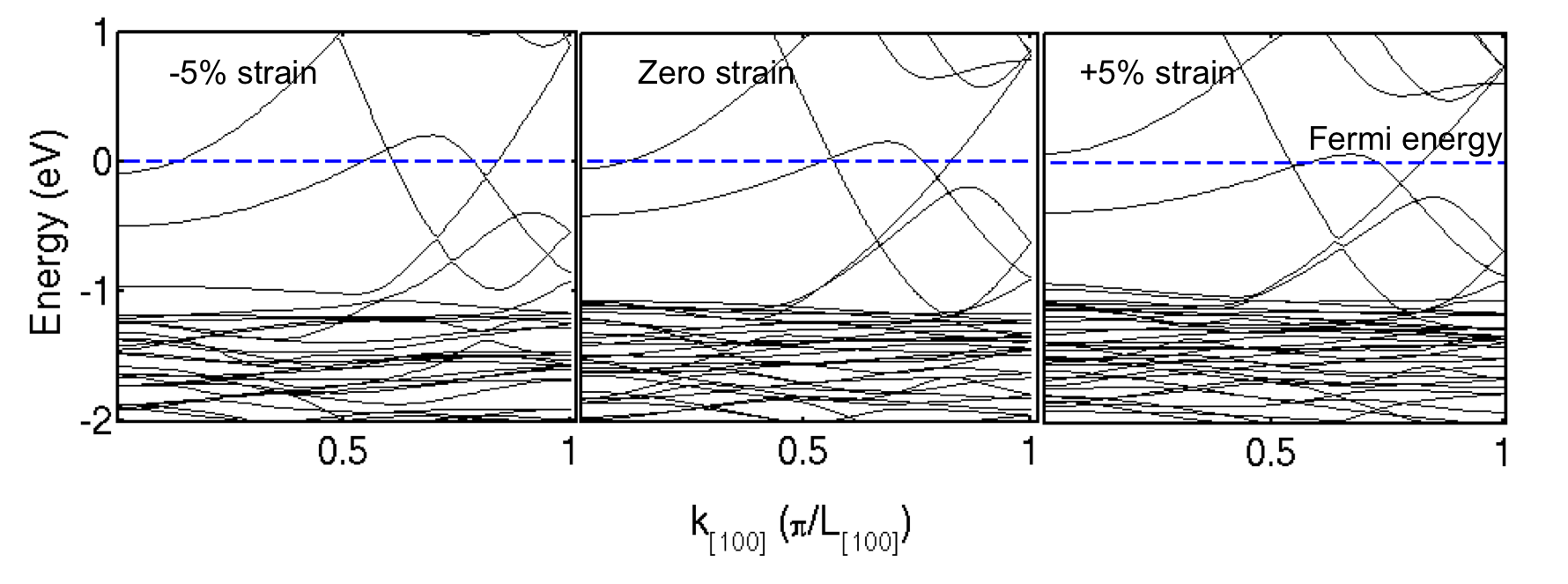}
	\caption{ Band structures of 2X2 unitcell cross section Cu [100] nanowires having (from left to right) at -5\%, 0 and +5\% homogeneous strain respectively.}
	\label{fig:100wirebandstructures}
\end{figure}

Figure \ref{fig:100wirebandstructures} further elaborates on this by comparing band structures of 2X2 unit cell Cu [100] nanowires under a variety of strain environments. Though overall band profile is similar on application of compressive and tensile strain, the cross sectional area changes significantly changing the density of conducting modes per unit cross sectional area.
%\item Not only does density of modes (DOM) per unit volume near FErmi Level increase, but so does DOS per unit volume. For e.g. Figure \ref{fig:Bulk_DOS_versus_strain} shows the effect of strain on bulk FCC Cu. DOS per unit volume also increases with application of compressive strain for same reason as DOM.
%\begin{figure}[H]
	%\centering
		%\includegraphics[width=3.5in]{Bulk_DOS_versus_strain.png}
	%\caption{ DOS versus strain in bulk FCC Cu}
	%\label{fig:Bulk_DOS_versus_strain}
%\end{figure}
%\item Nanowire ballistic conductance versus transport direction shows significant anisotropy. Fig \ref{fig:wiretransmissionversusarea}
%
%%(see Fig \ref{fig:CrossSectionsAll})
%\begin{figure} [here]
	%\centering
		%\includegraphics[width=4.0in]{wiretransmissionversusarea.pdf}
	%\caption{ Average Wire Transmission integrated versus nanowire cross sectional area for three different transport orientations. It is evident that there is significant anisotropy in ballistic conductance versus transport orientation}	
	%\label{fig:wiretransmissionversusarea}
%\end{figure}
\section{Conclusion}

In conclusion, a fully quantum mechanical study of the effect of quantum confinement, transport orientation and homogeneous strain on the ballistic conductance in Cu has been carried out using the new TB model. This study serves as a demonstration of the capabilities of the new TB model described in part I. It is also an initial step in a systematic, step-by-step theoretical investigation of factors that affect conductance of Cu. Additional effects, such as surface relaxation, phonon scattering, surface roughness, grain boundaries, alloying, aspect ratio, geometry and liner-interface bonding can now be added within the same TB framework to gauge the relative magnitude of the effect these factors have on Cu conductance.

It is found that quantum confinement results in significant conductance anisotropy versus transport orientation. The resistivity of [110] oriented nanowires seems least affected whereas [111] oriented nanowires show significant increase in resistivity with quantum confinement.
The physical mechanism behind the improvement of Cu conductance and reduction in resistivity with homogeneous compressive strain is clarified. In contrast with existing speculation in the literature \cite{haverty2007reducing}, it is found that compressive strain \textit{increases} the density of conducting modes per unit area thus improving Cu conductance and reducing Cu resistivity.

From a technological perspective, in addition to increasing grain size, engineering the grain profile of nanocrystalline Cu to ensure dominant [110] transport orientation may prove beneficial since this orientation shows least degradation in conductance due to quantum confinement. Using diffusion barrier metals and capping overlayers that induce compressive strain in Cu interconnects may improve Cu conductance and mitigate conductivity degradation of nanocrystalline Cu interconnects with decrease in interconnect dimensions.

We wish to re-iterate that the model and its exploration are fully implemented in the Nanoelectronic Modeling Suite NEMO5 \cite{fonseca2013efficient, steiger2011nanoelectronic}, which is released under an academic open source license on the nanoHUB \cite{NEMO5support}.

%\section{Discussion}
\begin{acknowledgments}
GH acknowledges helpful discussions with Sadasivan Shankar and Michael Haverty of Intel Corporation and Neerav Kharche at Brookhaven National Laboratory. This work was funded by Purdue University and the Semiconductor Research Corporation. The use of nanoHUB.org computational resources operated by the Network for Computational Nanotechnology funded by the US National Science Foundation under grants EEC-0228390, EEC-1227110, EEC-0228390, EEC-0634750, OCI-0438246, OCI-0721680 is gratefully acknowledged. The authors also gratefully acknowledge Kurt Stokbro and Anders Blom for the use of and help with the Atomistix Tool Kit (ATK).
\end{acknowledgments}

\bibliography{paper1}

\begin{thebibliography}{12}%
\makeatletter
\providecommand \@ifxundefined [1]{%
 \@ifx{#1\undefined}
}%
\providecommand \@ifnum [1]{%
 \ifnum #1\expandafter \@firstoftwo
 \else \expandafter \@secondoftwo
 \fi
}%
\providecommand \@ifx [1]{%
 \ifx #1\expandafter \@firstoftwo
 \else \expandafter \@secondoftwo
 \fi
}%
\providecommand \natexlab [1]{#1}%
\providecommand \enquote  [1]{``#1''}%
\providecommand \bibnamefont  [1]{#1}%
\providecommand \bibfnamefont [1]{#1}%
\providecommand \citenamefont [1]{#1}%
\providecommand \href@noop [0]{\@secondoftwo}%
\providecommand \href [0]{\begingroup \@sanitize@url \@href}%
\providecommand \@href[1]{\@@startlink{#1}\@@href}%
\providecommand \@@href[1]{\endgroup#1\@@endlink}%
\providecommand \@sanitize@url [0]{\catcode `\\12\catcode `\$12\catcode
  `\&12\catcode `\#12\catcode `\^12\catcode `\_12\catcode `\%12\relax}%
\providecommand \@@startlink[1]{}%
\providecommand \@@endlink[0]{}%
\providecommand \url  [0]{\begingroup\@sanitize@url \@url }%
\providecommand \@url [1]{\endgroup\@href {#1}{\urlprefix }}%
\providecommand \urlprefix  [0]{URL }%
\providecommand \Eprint [0]{\href }%
\providecommand \doibase [0]{http://dx.doi.org/}%
\providecommand \selectlanguage [0]{\@gobble}%
\providecommand \bibinfo  [0]{\@secondoftwo}%
\providecommand \bibfield  [0]{\@secondoftwo}%
\providecommand \translation [1]{[#1]}%
\providecommand \BibitemOpen [0]{}%
\providecommand \bibitemStop [0]{}%
\providecommand \bibitemNoStop [0]{.\EOS\space}%
\providecommand \EOS [0]{\spacefactor3000\relax}%
\providecommand \BibitemShut  [1]{\csname bibitem#1\endcsname}%
\let\auto@bib@innerbib\@empty
%</preamble>
\bibitem [{\citenamefont {Ke}\ \emph {et~al.}(2009)\citenamefont {Ke},
  \citenamefont {Zahid}, \citenamefont {Timoshevskii}, \citenamefont {Xia},
  \citenamefont {Gall},\ and\ \citenamefont {Guo}}]{PhysRevB.79.155406}%
  \BibitemOpen
  \bibfield  {author} {\bibinfo {author} {\bibfnamefont {Y.}~\bibnamefont
  {Ke}}, \bibinfo {author} {\bibfnamefont {F.}~\bibnamefont {Zahid}}, \bibinfo
  {author} {\bibfnamefont {V.}~\bibnamefont {Timoshevskii}}, \bibinfo {author}
  {\bibfnamefont {K.}~\bibnamefont {Xia}}, \bibinfo {author} {\bibfnamefont
  {D.}~\bibnamefont {Gall}}, \ and\ \bibinfo {author} {\bibfnamefont
  {H.}~\bibnamefont {Guo}},\ }\bibfield  {title} {\enquote {\bibinfo {title}
  {Resistivity of thin cu films with surface roughness},}\ }\href {\doibase
  10.1103/PhysRevB.79.155406} {\bibfield  {journal} {\bibinfo  {journal} {Phys.
  Rev. B}\ }\textbf {\bibinfo {volume} {79}},\ \bibinfo {pages} {155406}
  (\bibinfo {year} {2009})}\BibitemShut {NoStop}%
\bibitem [{\citenamefont {Zahid}\ \emph {et~al.}(2010)\citenamefont {Zahid},
  \citenamefont {Ke}, \citenamefont {Gall},\ and\ \citenamefont
  {Guo}}]{PhysRevB.81.045406}%
  \BibitemOpen
  \bibfield  {author} {\bibinfo {author} {\bibfnamefont {F.}~\bibnamefont
  {Zahid}}, \bibinfo {author} {\bibfnamefont {Y.}~\bibnamefont {Ke}}, \bibinfo
  {author} {\bibfnamefont {D.}~\bibnamefont {Gall}}, \ and\ \bibinfo {author}
  {\bibfnamefont {H.}~\bibnamefont {Guo}},\ }\bibfield  {title} {\enquote
  {\bibinfo {title} {Resistivity of thin cu films coated with ta, ti, ru, al,
  and pd barrier layers from first principles},}\ }\href {\doibase
  10.1103/PhysRevB.81.045406} {\bibfield  {journal} {\bibinfo  {journal} {Phys.
  Rev. B}\ }\textbf {\bibinfo {volume} {81}},\ \bibinfo {pages} {045406}
  (\bibinfo {year} {2010})}\BibitemShut {NoStop}%
\bibitem [{\citenamefont {Feldman}\ \emph {et~al.}(2010)\citenamefont
  {Feldman}, \citenamefont {Park}, \citenamefont {Haverty}, \citenamefont
  {Shankar},\ and\ \citenamefont {Dunham}}]{feldman2010simulation}%
  \BibitemOpen
  \bibfield  {author} {\bibinfo {author} {\bibfnamefont {B.}~\bibnamefont
  {Feldman}}, \bibinfo {author} {\bibfnamefont {S.}~\bibnamefont {Park}},
  \bibinfo {author} {\bibfnamefont {M.}~\bibnamefont {Haverty}}, \bibinfo
  {author} {\bibfnamefont {S.}~\bibnamefont {Shankar}}, \ and\ \bibinfo
  {author} {\bibfnamefont {S.~T.}\ \bibnamefont {Dunham}},\ }\bibfield  {title}
  {\enquote {\bibinfo {title} {Simulation of grain boundary effects on
  electronic transport in metals, and detailed causes of scattering},}\
  }\href@noop {} {\bibfield  {journal} {\bibinfo  {journal} {physica status
  solidi (b)}\ }\textbf {\bibinfo {volume} {247}},\ \bibinfo {pages}
  {1791--1796} (\bibinfo {year} {2010})}\BibitemShut {NoStop}%
\bibitem [{\citenamefont {Zhou}\ \emph {et~al.}(2008)\citenamefont {Zhou},
  \citenamefont {Sreekala}, \citenamefont {Ajayan},\ and\ \citenamefont
  {Nayak}}]{zhou2008resistance}%
  \BibitemOpen
  \bibfield  {author} {\bibinfo {author} {\bibfnamefont {Y.}~\bibnamefont
  {Zhou}}, \bibinfo {author} {\bibfnamefont {S.}~\bibnamefont {Sreekala}},
  \bibinfo {author} {\bibfnamefont {P.}~\bibnamefont {Ajayan}}, \ and\ \bibinfo
  {author} {\bibfnamefont {S.}~\bibnamefont {Nayak}},\ }\bibfield  {title}
  {\enquote {\bibinfo {title} {Resistance of copper nanowires and comparison
  with carbon nanotube bundles for interconnect applications using first
  principles calculations},}\ }\href@noop {} {\bibfield  {journal} {\bibinfo
  {journal} {Journal of Physics: Condensed Matter}\ }\textbf {\bibinfo {volume}
  {20}},\ \bibinfo {pages} {095209} (\bibinfo {year} {2008})}\BibitemShut
  {NoStop}%
\bibitem [{\citenamefont {Kharche}\ \emph {et~al.}(2011)\citenamefont
  {Kharche}, \citenamefont {Manjari}, \citenamefont {Zhou}, \citenamefont
  {Geer},\ and\ \citenamefont {Nayak}}]{kharche2011comparative}%
  \BibitemOpen
  \bibfield  {author} {\bibinfo {author} {\bibfnamefont {N.}~\bibnamefont
  {Kharche}}, \bibinfo {author} {\bibfnamefont {S.~R.}\ \bibnamefont
  {Manjari}}, \bibinfo {author} {\bibfnamefont {Y.}~\bibnamefont {Zhou}},
  \bibinfo {author} {\bibfnamefont {R.~E.}\ \bibnamefont {Geer}}, \ and\
  \bibinfo {author} {\bibfnamefont {S.~K.}\ \bibnamefont {Nayak}},\ }\bibfield
  {title} {\enquote {\bibinfo {title} {A comparative study of quantum transport
  properties of silver and copper nanowires using first principles
  calculations},}\ }\href@noop {} {\bibfield  {journal} {\bibinfo  {journal}
  {Journal of Physics: Condensed Matter}\ }\textbf {\bibinfo {volume} {23}},\
  \bibinfo {pages} {085501} (\bibinfo {year} {2011})}\BibitemShut {NoStop}%
\bibitem [{\citenamefont {Haverty}\ \emph {et~al.}(2007)\citenamefont
  {Haverty}, \citenamefont {Shankar}, \citenamefont {O'brien},\ and\
  \citenamefont {Park}}]{haverty2007reducing}%
  \BibitemOpen
  \bibfield  {author} {\bibinfo {author} {\bibfnamefont {M.}~\bibnamefont
  {Haverty}}, \bibinfo {author} {\bibfnamefont {S.}~\bibnamefont {Shankar}},
  \bibinfo {author} {\bibfnamefont {K.}~\bibnamefont {O'brien}}, \ and\
  \bibinfo {author} {\bibfnamefont {S.}~\bibnamefont {Park}},\ }\href@noop {}
  {\enquote {\bibinfo {title} {Reducing resistivity in metal interconnects by
  compressive straining},}\ } (\bibinfo {year} {2007}),\ \bibinfo {note} {uS
  Patent App. 11/771,476}\BibitemShut {NoStop}%
\bibitem [{\citenamefont {Fonseca}\ \emph {et~al.}(2013)\citenamefont
  {Fonseca}, \citenamefont {Kubis}, \citenamefont {Povolotskyi}, \citenamefont
  {Novakovic}, \citenamefont {Ajoy}, \citenamefont {Hegde}, \citenamefont
  {Ilatikhameneh}, \citenamefont {Jiang}, \citenamefont {Sengupta},
  \citenamefont {Tan} \emph {et~al.}}]{fonseca2013efficient}%
  \BibitemOpen
  \bibfield  {author} {\bibinfo {author} {\bibfnamefont {J.}~\bibnamefont
  {Fonseca}}, \bibinfo {author} {\bibfnamefont {T.}~\bibnamefont {Kubis}},
  \bibinfo {author} {\bibfnamefont {M.}~\bibnamefont {Povolotskyi}}, \bibinfo
  {author} {\bibfnamefont {B.}~\bibnamefont {Novakovic}}, \bibinfo {author}
  {\bibfnamefont {A.}~\bibnamefont {Ajoy}}, \bibinfo {author} {\bibfnamefont
  {G.}~\bibnamefont {Hegde}}, \bibinfo {author} {\bibfnamefont
  {H.}~\bibnamefont {Ilatikhameneh}}, \bibinfo {author} {\bibfnamefont
  {Z.}~\bibnamefont {Jiang}}, \bibinfo {author} {\bibfnamefont
  {P.}~\bibnamefont {Sengupta}}, \bibinfo {author} {\bibfnamefont
  {Y.}~\bibnamefont {Tan}},  \emph {et~al.},\ }\bibfield  {title} {\enquote
  {\bibinfo {title} {Efficient and realistic device modeling from atomic detail
  to the nanoscale},}\ }\href@noop {} {\bibfield  {journal} {\bibinfo
  {journal} {Journal of Computational Electronics}\ ,\ \bibinfo {pages} {1--9}}
  (\bibinfo {year} {2013})}\BibitemShut {NoStop}%
\bibitem [{\citenamefont {Steiger}\ \emph {et~al.}(2011)\citenamefont
  {Steiger}, \citenamefont {Povolotskyi}, \citenamefont {Park}, \citenamefont
  {Kubis}, \citenamefont {Hegde}, \citenamefont {Haley}, \citenamefont
  {Rodwell},\ and\ \citenamefont {Klimeck}}]{steiger2011nanoelectronic}%
  \BibitemOpen
  \bibfield  {author} {\bibinfo {author} {\bibfnamefont {S.}~\bibnamefont
  {Steiger}}, \bibinfo {author} {\bibfnamefont {M.}~\bibnamefont
  {Povolotskyi}}, \bibinfo {author} {\bibfnamefont {H.}~\bibnamefont {Park}},
  \bibinfo {author} {\bibfnamefont {T.}~\bibnamefont {Kubis}}, \bibinfo
  {author} {\bibfnamefont {G.}~\bibnamefont {Hegde}}, \bibinfo {author}
  {\bibfnamefont {B.}~\bibnamefont {Haley}}, \bibinfo {author} {\bibfnamefont
  {M.}~\bibnamefont {Rodwell}}, \ and\ \bibinfo {author} {\bibfnamefont
  {G.}~\bibnamefont {Klimeck}},\ }\bibfield  {title} {\enquote {\bibinfo
  {title} {The nanoelectronic modeling tool nemo 5: Capabilities, validation,
  and application to sb-heterostructures},}\ }in\ \href@noop {} {\emph
  {\bibinfo {booktitle} {Device Research Conference (DRC), 2011 69th Annual}}}\
  (\bibinfo {organization} {IEEE},\ \bibinfo {year} {2011})\ pp.\ \bibinfo
  {pages} {23--26}\BibitemShut {NoStop}%
\bibitem [{\citenamefont {Datta}(1997)}]{datta1997electronic}%
  \BibitemOpen
  \bibfield  {author} {\bibinfo {author} {\bibfnamefont {S.}~\bibnamefont
  {Datta}},\ }\href@noop {} {\emph {\bibinfo {title} {Electronic transport in
  mesoscopic systems}}}\ (\bibinfo  {publisher} {Cambridge university press},\
  \bibinfo {year} {1997})\BibitemShut {NoStop}%
\bibitem [{\citenamefont {Lundstrom}(2009)}]{lundstrom2009fundamentals}%
  \BibitemOpen
  \bibfield  {author} {\bibinfo {author} {\bibfnamefont {M.}~\bibnamefont
  {Lundstrom}},\ }\href@noop {} {\emph {\bibinfo {title} {Fundamentals of
  carrier transport}}}\ (\bibinfo  {publisher} {Cambridge University Press},\
  \bibinfo {year} {2009})\BibitemShut {NoStop}%
\bibitem [{\citenamefont {Ashcroft}\ and\ \citenamefont
  {Mermin}(1976)}]{ashcroft1976solid}%
  \BibitemOpen
  \bibfield  {author} {\bibinfo {author} {\bibfnamefont {N.~W.}\ \bibnamefont
  {Ashcroft}}\ and\ \bibinfo {author} {\bibfnamefont {N.~D.}\ \bibnamefont
  {Mermin}},\ }\bibfield  {title} {\enquote {\bibinfo {title} {Solid state
  physics (holt},}\ }\href@noop {} {\bibfield  {journal} {\bibinfo  {journal}
  {Rinehart and Winston, New York}\ }\textbf {\bibinfo {volume} {19761}}
  (\bibinfo {year} {1976})}\BibitemShut {NoStop}%
\bibitem [{NEM(2013)}]{NEMO5support}%
  \BibitemOpen
  \href {https://nanohub.org/groups/nemo5distribution} {\enquote {\bibinfo
  {title} {Nemo5 support and distribution group on the nanohub},}\ } (\bibinfo
  {year} {2013})\BibitemShut {NoStop}%
\end{thebibliography}%

\end{document}